\documentstyle[A4]{article}

\def\w{\omega}
\def\beq{\begin{equation}}
\def\eeq{\end{equation}}
\def \a{\alpha}
\def \b{\beta}

\def \e{\epsilon}

\newtheorem{prop}{Proposition}

\begin{document}

\begin{titlepage}

\title{Affine Toda field theory as a 3-dimensional integrable
system}

\author{R.M. Kashaev\thanks{On leave of absence from
St. Petersburg Branch of the Steklov Mathematical Institute,
Fontanka 27, St. Petersburg 191011, RUSSIA}\\ \\
Laboratoire de Physique Th\'eorique
{\sc enslapp}\thanks{URA 14-36 du CNRS,
associ\'ee \`a l'E.N.S. de Lyon,
et \`a l'Universit\`e de Savoie}
\\
ENSLyon,
46 All\'ee d'Italie,\\
69007 Lyon, FRANCE\\
\\
N. Reshetikhin\thanks{The work of N.R. was partially
supported by NSF Grant DMS-9692-120}\\ \\
Department of Mathematics,\\ University of California, \\
Berkeley, CA 94720, U.S.A.\\}

\date{June 1995}

\maketitle

\abstract{The affine Toda field theory is studied as a 2+1-dimensional system.
The third dimension appears as the discrete space dimension, corresponding
to the simple roots in the $A_N$ affine root system, enumerated according to
the cyclic order on the  $A_N$ affine Dynkin diagram. We show that
there exists a natural discretization of the affine Toda theory,
where the equations of motion are invariant with respect to
permutations of all discrete coordinates. The discrete evolution operator
is constructed explicitly. The thermodynamic Bethe ansatz of
the affine Toda system is studied in the
limit $L,N\to\infty$. Some conjectures about the structure of
the spectrum of the corresponding discrete models are stated.}
\vskip 2cm

\rightline{{\small E}N{\large S}{\Large L}{\large A}P{\small P}-L-xxx/95}

\end{titlepage}

\section{Introduction}
The Toda field theories were extensively studied as classical and
quantum integrable field theories.
The latest developments in the
study of these models are described in the paper \cite{Corr}.
In the quantum case Toda field theories provide examples of
integrable models of quantum field theories with a scalar
factorizable $S$-matrix. The affine Toda field theory of $A_N$ type
describes $N$ scalar fields interacting nonlinearly with the
Lagrangian
\begin{equation}  \label{Toda}
L_{AT}=\int \sum_{i=1}^N \left({1\over2}
({\partial \phi_i \over \partial t})^2-
{1\over2}({\partial \phi_i \over \partial x})^2
-{{M^2}\over {\beta}^2}\exp(\beta(\phi_i-\phi_{i+1}))\right) dx.
\end{equation}
Here it is assumed that $\phi_{N+1}=\phi_1$.

The mass spectrum and the scattering amplitudes
of this model of the quantum field theory
were suggested in \cite{AFZ} on the base of the bootstrap
principle and the perturbation theory.

The spectrum of the model consists of $N-1$ massive particles
with masses $M_l=M\sin({{{\pi}l}/{N}})$, where $M$ is the
renormalized mass and $l=1,\ldots, N-1$. The scattering amplitude of
the $l$-th particle on the $k$-th particle is given by the following product
\begin{equation} \label{S-matr}
S_{k,l}=\prod_{{1\le n\le k}\ {1\le m\le l}}S_{11}
\left(\theta+i{\pi(k-1-2n)\over N}-i{\pi (l-1-2m)\over N}\right),
\end{equation}
where
$$
S_{11}(\theta)={{\sinh({{\theta}\over{2}}+{i\pi\over N})
\sinh({\theta\over 2}-{i\pi\over N}+{ib\over 2})
\sinh({\theta\over 2}-{ib\over2})}\over{\sinh({\theta\over 2}-{i\pi\over N})
\sinh({\theta\over 2}+{i\pi\over N}-{ib\over 2})
\sinh({\theta\over 2}+{ib\over 2})}}.
$$
It is interesting to compare the Toda model (\ref{Toda}) with the
so called open Toda field theory, described by the same Lagrangian
(\ref{Toda}), but without the periodicity condition $\phi_{N+1}=\phi_1$.
The Liouville theory \cite{Polyak}, which describes the
2D-gravity, is a simplest example of such model with $N=1$.
The periodic and open Toda field theories have completely
different structure of dynamics for finite $N$ \cite{SavLezn,LS}. In
particular,
the open Toda chain has massless spectrum.
Moreover, the open  Toda model (with $N$ fields) can be obtained from the
model (\ref{Toda}) in the limit $Z\to 0$, $M=mZ$,
$\phi_i=\tilde\phi_i+(i2/\b)\ln Z$
for $i=1,\dots,N$, and $m$ being fixed, where the Lagrangian (\ref{Toda})
becomes the Lagrangian of the open Toda field theory
with the mass parameter $m$ and with the fields $\tilde\phi_i$,
$i=1,\ldots,N$.

In this paper we investigate the possibility of interpreting the
affine Toda field theory as the model in $2+1$-dimensions
with the discrete second space coordinate $i=1,\dots,N$.
In this interpretation the model (\ref{Toda})
corresponds to the periodic boundary conditions in the second space
dimension, while the open Toda model corresponds to the open
boundary conditions.
Notice, that if we would not have the first space coordinate
$x$, then the interpretation
of the components of the field $\{\phi_i\}$
as an extra space dimension would correspond to the original
physical model, suggested by Toda for the description of long molecules
\cite{Todabook}.

The following results came out of the study of such interpretation
of the Toda field theory:
\begin{itemize}
\item The discrete local integrable version of the affine Toda
        field theory is proposed. The equations of motion for certain
        variables  in this discrete
        model are invariant with respect to any permutation of coordinates.
        In the classical limit these variables are the $\tau$
         functions \cite{Hirota,JiMi}. When the space-time
        lattice is finite
        the model provides a finite dimensional approximation to the
        Toda field theory.

\item It is shown that in the limit $N\to\infty$ the system has a
      thermodynamical limit. The spectrum of the resulting theory consists
      of massless scalar particles. The density of the free energy
      is computed using the thermodynamical Bethe ansatz.
\end{itemize}

The paper is organized as follows. In section~\ref{sec1} we describe the
discrete integrable version of the Toda field theory and show that in the
classical and continuous limit it is reduced to the classical Toda field theory
(\ref{Toda}). In the next section, we construct explicitly the evolution
operator and, using the
discrete Lax operators \cite{Discr}, describe the integrals of motion. The
construction is very similar to
the one, used for the discrete Sine-Gordon model, see
\cite{FaddVolk,BBR,BPK}. In section~\ref{sec3}
we study this discrete system as a 3-dimensional discrete field theory.
In appropriate variables, this provides a system of difference
equations for the operators in the extended algebra of observables, which is
invariant with respect to permutations of coordinates. In the classical
case this system first appeared in \cite{Hirota,JiMi}.
Section~\ref{sec4} contains the analysis of the thermodynamical limit of the
affine Toda system, regarded as a 3-dimensional field theory. It is shown
that there exists a 3-dimensional thermodynamical limit in which the
excitations
are massless scalar particles.

\section{The discrete space-time Toda field theory}\label{sec1}
\setcounter{equation}{0}

\subsection{} The classical Toda field theories in the
continuous space-time are infinite dimensional integrable
Hamiltonian systems.
The Lax representation for these systems is well known  \cite{SavLezn}.
In this case
the Lax representation corresponds to certain choice of
the symplectic leaf in the appropriate Lie bialgebra
\cite{Kost,RSTS}. This is equivalent to the
fact that the Poisson brackets between matrix elements of
the Lax operators are given by the so called $r$-matrix Poisson
brackets \cite{FadTakht}.

If one wants to find an integrable discrete space-time analog of the
Toda field theory, the natural way to proceed is to find the
appropriate discrete Lax pair \cite{Discr}. A discrete Hamiltonian
system with the Lax representation is usually integrable if the Lax
operator has the $r$-matrix Poisson brackets.
In more geometrical language this
means that the Lax operator describes certain symplectic leaf
in the Poisson Lie group $\widehat{SL}(N)$. The factorization
map, constructed in \cite{MoVe,Deift,Resh} provides a Poisson
map from the Poisson Lie group to itself, which, being restricted to
the corresponding symplectic leaf, gives the evolution
with equations of motion for the coordinate functions,
given by the discrete Lax equation.

The discretization on the classical level reduces the infinite dimensional
integrable Hamiltonian system to the finite dimensional one. The next step is
the quantization.

The quantization replaces the Poisson evolution map by an automorphism
of the quantum algebra of observables. Since the classical
system is integrable we certainly want to construct
an integrable quantization (with extra integrals of motion). If
the quantum system can be described in terms of differential
operators on an $n$-dimensional manifold, there should be $n$ such integrals.
It is known that if the system admits the quantum Lax representation, and
if the quantum Lax operators have the so called $R$-matrix commutation
 relations,
then one can construct the integrals of such evolution, using
the traces of appropriate products of quantum Lax operators.

In the algebraic language, the quantum Lax operator is the universal
$R$-matrix of certain factorizable Hopf algebra evaluated on the tensor
product of two representations. For the details see \cite{Resh}.

Now we will describe the quantum discrete Toda system and then show
that in the classical limit it gives the discrete version of
the Toda field theory. The classical equations of motion for
the discrete Toda field theory first appeared in \cite{Hirota}
as bilinear difference equations for the corresponding $\tau$
function. The Lax representation for these equations is
described in \cite{Ward} without the Hamiltonian interpretation.
The Lax representation together with the Hamiltonian interpretation
for the Toda chain with discrete time was found in \cite{Suris}.

\subsection{} Here we fix some notations, used throughout the paper.

A parameter $q$ is some non-zero complex number,
$h\equiv\ln(q);$
$L$ and $N$ are arbitrary positive integers,
$F\equiv(L,N)$, the greatest common divisor of $L$ and $N$, and $J=LN/F$,
the smallest common multiple.

For any positive integer $m$ and any integer $n$ we
denote by $[n]_m$ the non-negative integer, such that
\beq\label{[n]}
[n]_m=n\pmod m,\quad 0\le[n]_m<m.
\eeq
For example, we write $[x]_m=0$ instead of $ x=0\pmod m$.
The function
\beq\label{eps_m}
\varepsilon_m(n)=\frac{m-1-2[n]_m}{m},
\eeq
is periodic in argument $n$:\[\varepsilon_m(n+m)=\varepsilon_m(n).\]
It also satisfies the following relations:
\[\varepsilon_m(n)+\varepsilon_m(-1-n)=0,\]
\[
\varepsilon_m(n)-\varepsilon_m(n-1)=2(\delta_m(n)-1/m),\]
where
\beq {\delta}_m(n)=\left\{ \begin{array}{ll}
1 ,& \mbox{if }[n]_m=0; \\
0, & \mbox{otherwise.}
\end{array}
\right.
\eeq
Notice, that for any finite $n$
\[\lim_{m\rightarrow\infty}\varepsilon_m(n)=\varepsilon(n)\equiv\left\{
\begin{array}{rl}
                                           1,&\mbox{if }n\ge0;\\
                                           -1,&\mbox{otherwise.}
                                          \end{array}
\right.\]

\subsection {}Define the algebra $C_q(N,L)$,
generated by invertible elements \[\chi_i(n),\quad i=1,\ldots,N;\quad
n=1,\ldots, 2L,\] with the following
commutation relations
\[
\chi_i(n)\chi_j(m)=\chi_j(m)\chi_i(n),\quad [n-m]_2=0;\]
\beq\chi_i(n)\chi_j(m)=q^{\Omega(i-j,(n-m-1)/2)}
\chi_j(m)\chi_i(n),\quad [m]_2-[n]_2=1,
\eeq
where
\beq \label{omeg}
\Omega(i,n)=2\delta_L(n)\left(\delta_N(i)-\delta_N(i+1)\right)
+2\delta_L(n+1)\left(\delta_N(i)-\delta_N(i-1)\right).
\eeq
Consider two maps $\kappa_{\pm}\colon C_q(N,L)\rightarrow C_q(N,L)$ defined on
the generators $\chi_i(n)$ as follows:
\[
\kappa_\pm\colon\chi_i(n\mp1)\mapsto\chi_i(n),\quad[n]_2=0;\]
\beq\label{aut}
\kappa_\pm\colon\chi_i(n\mp1)\mapsto
q^{-2}\frac{1-\chi_{i-1}(n+1)}{1-q^{-2}
\chi_i^{-1}(n+1)}\
\frac{1-\chi_{i+1}(n-1)}{1-q^{-2}\chi_i^{-1}(n-1)}\ \chi_i^{-1}
(n),\quad [n]_2=1.
\eeq
Here and below we assume that the subindex $i$, enumerating generators
$\chi_i(n)$, and their argument $n$ are taken modulo $N$ and $2L$,
respectively:
\[\chi_{i+N}(n)=\chi_i(n+2L)=\chi_i(n).\]
One can also say that $\chi_i(n)$ satisfy the
`periodic boundary conditions' in $i$ and $n$.
\begin{prop}
The maps $\kappa_\pm$ in (\ref{aut}), extended by linearity to the whole
algebra\footnote{Strictly
speaking, the maps $\kappa_{\pm}$ act from the algebra to
its completion, which can be defined in many ways. For
example, one can consider Laurent power series in $\chi$, or $\chi-1$, or
one can realize $\chi$ as operators in a Hilbert space
and use the spectral theorem for the definition of
rational functions of an operator. We will ignore these
problems when possible.} $C_q(N,L)$, determine automorphisms
of this algebra.
\end{prop}

For each $i=1,\ldots,N$, $n=1,\ldots, 2L$, define the `trajectory' of
the discrete Toda system as the sequence of elements in  $C_q(N,L)$:
\beq
\chi_i(n{\pm}1,t+1)=\kappa_{\pm}(\chi_i(n,t)); \quad
 \chi_i(n)=\left\{\begin{array}{ll}
\chi_i(n,1), &\mbox{if }[n]_2=0;\\
\chi_i(n,0),&\mbox{otherwise.}
\end{array}\right.
\eeq
Notice that the fields $\chi_i(n,t)$ are defined only in the case
$[n+t]_2=1$. They satisfy the following equations of motions
\begin{eqnarray}
\label{toda}
\lefteqn{q^{-2}\chi_i(n,t+1)\ \chi_i(n,t-1)
=\chi_i(n+1,t)\ \chi_i(n-1,t)}\qquad\qquad\qquad\nonumber   \\
&\times&\frac{1-\chi_{i-1}(n+1,t)}{1-q^2\chi_i(n+1,t)}\
\frac{1-\chi_{i+1}(n-1,t)}{1-q^2\chi_i(n-1,t)}.
\end{eqnarray}
These are the discrete space-time Heisenberg equations of motion for the
discrete Toda field theory.

Now, let us show how to recover the classical continuous Toda field theory
(\ref{Toda})
from this system. First consider the classical discrete Toda, which corresponds
to the limit $q\to 1$ in the quantum discrete Toda model
described above. Let $\varepsilon$ be the lattice spacing both in the time and
space directions. Define the new field:
\beq\label{varphi}
\varphi_i(n\varepsilon,t\varepsilon)=\frac{1}{\b}\ln\left(-\frac{\chi_i(n,t)}
{(\varepsilon M)^2}\right).
\eeq
Substituting this into (\ref{toda}), we obtain ($q=1$):
\[
\exp\left(\b(\varphi_i(x,t+\varepsilon)+\varphi_i(x,t-\varepsilon)
-\varphi_i(x+\varepsilon,t)-\varphi_i(x-\varepsilon,t))\right)\]
\[=
\frac{1+(\varepsilon M)^2\exp(\b\varphi_{i-1}(x+\varepsilon,t))}{1+(\varepsilon
M)^2
\exp(\b\varphi_i(x+\varepsilon,t))}\
\frac{1+(\varepsilon M)^2\exp(\b\varphi_{i+1}(x-\varepsilon,t))}{1+(\varepsilon
M)^2
\exp(\b\varphi_i(x-\varepsilon,t))}.
\]
In the limit $\varepsilon\rightarrow0$ these equations  reduce to
\[
\frac{\partial^2\varphi_i}{\partial t^2}
-\frac{\partial^2\varphi_i}{\partial
x^2}=\frac{M^2}{\b}\left(\exp(\b\varphi_{i+1})
+\exp(\b\varphi_{i-1})-2\exp(\b\varphi_i)\right),
\]
which coincide with the Euler-Lagrange equations for the system (\ref{Toda}),
written for the fields
\beq\label{corresp}\varphi_i=\phi_i-\phi_{i+1}.\eeq

\section{Evolution operator and integrals of motion}\label{sec2}
\setcounter{equation}{0}

In this section first we investigate the nature of the automorphisms
$\kappa_{\pm}$ defined in (\ref{aut}). In particular we will see how to realize
them
as inner automorphisms of some completion of $C_q(N,L)$.
 Then we construct the elements
of the algebra of observables, which are conserved under the evolution
(\ref{toda}).

\subsection{} Let us complete the algebra $C_q(N,L)$ by formal power
series in $\chi_i(n)-1$. The result denote as $C'_q(N,L)$.
Introduce the elements
$\psi_i(n)$, exponentials of which coincide with $\chi_i(n)$:
\beq
\chi_i(n)=-(\varepsilon M)^2\exp(2\psi_i(n)),\quad i=1,\ldots,N;\quad
n=1,\ldots,2L.
\eeq
In fact, the operators $\psi_i(n)$, up to a multiplicative factor and a
redefinition of the arguments, are the quantum analogs of the fields
(\ref{varphi}) at a fixed time.

The following commutation relations hold:
\[
[\psi_i(n),\psi_j(m)] = \frac{h}{4}\Omega\left(i-j,\frac{n-m-1}{2}\right),\quad
[m]_2-[n]_2=1;\]
\[
[ \psi_i(n),\psi_j(m) ] =  0,\quad [m]_2=[n]_2,
\]
where $\Omega(i,n)$ is given in (\ref{omeg}).

The center of this algebra is generated by the elements, corresponding
to the eigenvectors of the matrix:
\[
\left[\begin{array}{cc}
0&\Omega(i-j,n-m)\\
-\Omega(j-i,m-n)&0\end{array}\right]
\]
with zero eigenvalues.
We will use the following part of it.
\begin{prop}
The following elements\footnote{The
elements $u(n)$, $v(n)$ are not independent, for example,
it is easy to see that
$\sum_{n=1}^Lu(2n)=\sum_{k=1}^Fv(2k),\quad
\sum_{n=1}^Lu(2n-1)=\sum_{k=1}^Fv(2k-1).$}
lie in the center of  $C'_q(N,L)$:
\[
u(n)\equiv\sum_{i=1}^N\psi_i(n),\quad n=1,\ldots,2L;\]
\beq \label{center}
v(2k-(1\pm1)/2)\equiv\sum_{i=1}^N\sum_{n=1}^L\psi_i(2n-(1\pm1)/2)
\delta_F(i+n-k),\quad k=1,\ldots,F.
\eeq
\end{prop}
Consider the quotient algebra of $C'_q(N,L)$ with the extra relations
\beq \label{charge}
u(n)=v(k)=0,\quad n=1,\ldots,2L;\ k=1,\ldots,2F.
\eeq
The automorphisms $\kappa_\pm$ can be represented as the compositions
\beq\label{compos}
\kappa_\pm=\sigma\circ\lambda_\pm,
\eeq
where
\beq
\lambda_\pm(\chi_i(n))=\left\{\begin{array}{ll}
                              (\varepsilon M)^4\chi_i^{-1}(n\pm1),&\mbox{if }
[n]_2=0;\\
                               \chi_i(n\pm1),&\mbox{otherwise;}
\end{array}\right.
\eeq
and
\[
\sigma(\chi_i(n))=\chi_i(n),\quad [n]_2=0;\]
\beq\label{sigma} \sigma(\chi_i(n))=
                             q^2(\varepsilon
M)^4\chi_i(n)\frac{1-q^{-2}\chi_i^{-1}(n+1)}{1-\chi_{i-1}(n+1)}\
\frac{1-q^{-2}\chi_i^{-1}(n-1)}{1-\chi_{i+1}(n-1)},\quad [n]_2=1.
\eeq
To describe explicitly the evolution operator, introduce some auxiliary
objects.

Define the function
\beq\label{gxy}
g(x,y)=\frac{1}{2}\sum_{j=1}^{N/F}
\varepsilon_J\left(\frac{x-y-1}{2}+jL\right)
\varepsilon_N\left(\frac{x+y-1}{2}+jL\right),\quad [x+y]_2=1.
\eeq
It is symmetric under the reflection
\[
g(x,y)=g(-x,-y),\] and satisfies the difference equation:
\begin{eqnarray*}
\lefteqn{g(x+1,y)+g(x-1,y)-g(x,y+1)-g(x,y-1)}\\
&=&2\left(\delta_L\left(\frac{x-y}{2}\right)(\delta_N(y)
-N^{-1})-J^{-1}\delta_F\left(\frac{x+y}{2}\right)+(LN)^{-1}\right).
\end{eqnarray*}
Consider the quadratic form
\beq
H(\psi)=\frac{1}{2}
\sum_{i,j=1}^N\sum_{m,n=1}^Ls(i-j,m-n)\psi_i(2m)\psi_j(2n),
\eeq
where
\beq
s(i,n)=g(i+2n-1,i)+g(i+2n+1,i),
\eeq
and the following element in $C'_q(N,L)$:
\beq
U=\exp(H(\psi)/h)\prod_{i=1}^N\prod_{n=1}^L\Psi(\chi_i(2n)),
\eeq
with $\Psi(x)$ being some solution to the functional equation:
\beq\label{dilog}
\Psi(xq^{-2})=(1-x)\Psi(x).
\eeq
For example, if $|q|<1$, the solution to (\ref{dilog}), regular at $x=0$, has
the form
\[\Psi(x)=(q^2x;q^2)_\infty\equiv\prod_{n=1}^\infty(1-xq^{2n}).\]
\begin{prop}
The automorphism $\sigma$ (\ref{sigma}) is inner one in $C'_q(N,L)$ and is
represented by the element $U$:
\[ U^{-1}\chi_i(n)U=\sigma(\chi_i(n)),
\]
provided the central elements (\ref{center}) are chosen as in (\ref{charge}).
\end{prop}
Notice, that the restrictions (\ref{charge}) are not essential. In the case
the central elements are given arbitrary numerical values, the forms of
automorphisms $\lambda_\pm$ and $\sigma$ should be simultaneously changed by
appropriate numerical factors, keeping the compositions (\ref{compos}) fixed.

The automorphisms $\kappa_\pm$ correspond to time evolution combined with space
translations, while their composition
\beq\label{kappa}
 \kappa\equiv\kappa_+\circ\kappa_-=\kappa_-\circ\kappa_+,\eeq
corresponds to pure time translation:
\beq\label{2step}
\chi_i(n,t+2)=\kappa(\chi_i(n,t))=(VU)^{-1}\lambda(\chi_i(n,t))VU
=(\varepsilon M)^4(VU)^{-1}\chi_i^{-1}(n,t)VU,
\eeq
where
\[ V\equiv\lambda_+(U)=\lambda_-(U),\]
\[ \lambda\equiv\lambda_+\circ\lambda_-=\lambda_-\circ\lambda_+.\]
Note, that automorphisms $\lambda_\pm$ themselves are also inner ones in
$C'_q(N,L)$, but their role is auxiliary in the sense that in the `two-step'
time evolution (\ref{2step}) they combine into the involution $\lambda$.

\subsection {}Here we describe some algebra which will be used
in the description of quantum integrals for the evolution (\ref{toda}).

 Denote by $A_q(N)$ the algebra,
generated by invertible elements $a_i, b_i$, $i=1,\dots,N$,
with the following determining relations:
\beq  \label{a,b-alg}
a_i b_{i-1}=q b_{i-1} a_i, \quad
a_i b_{i}=q^{-1} b_{i} a_i.
\eeq
All other generators $a$ and $b$ commute.

Consider the following elements in
$\mbox{End}(C^N)\otimes A_q(N)$
\begin{equation}  \label{Lplus}
L^+(z)=\sum_{1\le i\le N}e_{i,i}\otimes a_i+
\sum_{1\le i< N}e_{i,i+1}\otimes b_i+z^{-2}e_{N,1}\otimes b_N,
\end{equation}
\beq        \label{Lmin}
L^-(z)=
\sum_{1\le i\le N}e_{i,i}\otimes a_i^{-1}+\sum_{1\le i< N}e_{i+1,i}\otimes
b_i+z^{2}e_{1,N}\otimes b_N.
\eeq

It is not difficult to check that these elements satisfy the following
identities in $\mbox{End}(C^N)\otimes \mbox{End}(C^N)\otimes A_q(N)$:
\begin{eqnarray}\label{rll}
R(x/y)L^+(x)\otimes L^+(y)&=&
(1\otimes L^+(x))(L^+(y)\otimes 1)R(x/y),\nonumber\\
L^-(x)\otimes L^-(y)R(x/y)&=&
R(x/y)(1\otimes L^-(y))(L^-(x)\otimes 1).
\end{eqnarray}
Here the tensor product is taken over the algebra $A_q(N)$ (tensor product of
matrices where the elements are multiplied in $A_q(N)$). The matrix
$R(z)$ is $N^2\times N^2$ matrix acts trivially in  $A_q(N)$ and
is the `fundamental' $U_q(\widehat {sl}(N))$ $R$-matrix :

\begin{eqnarray}
\lefteqn{R(z)=(qz-q^{-1}z^{-1})
\sum_{1\le i\le N}e_{i,i}\otimes e_{i,i}
+(z-z^{-1})\sum_{1\le i\ne j\le N}e_{i,i}\otimes
e_{j,j}}\qquad\qquad\nonumber\\
&+&(q-q^{-1})\sum_{1\le i\ne j\le N} (ze_{i,j}\otimes e_{j,i}+z^{-1}
ze_{j,i}\otimes e_{i,j}).
\end{eqnarray}

In terms of quantized universal enveloping algebras the matrices
$L^{\pm}$ describe the `minimal' representations \cite{JM,BKMS} of the
quantized Borel subalgebra $U_q(\widehat b_+)$ of the
quantized universal enveloping algebra of $\widehat {sl}(N)$.

\subsection{} Here we construct integrals and the quantum Lax pair for
the quantum discrete complexified Toda
field theory of affine $A_N$ type.

Consider the algebra $A_q(N,L)=A_q(N)^{\otimes 2L}$. It is
is generated by the elements
\beq\label{anbn} a_i(n)=1\otimes\dots \otimes a_i\otimes
\dots \otimes 1, \quad b_i(n)=1\otimes\dots \otimes b_i\otimes
\dots \otimes 1, \eeq where $n=1,\ldots,2L$.
Define the elements $L_n^{\pm}(z)$ in $\mbox{End}(C^N)\otimes A_q(N,L)$
\[
L_{n}^+(z)=\sum_{1\le i\le N}e_{i,i}\otimes a_i(n)+
\sum_{1\le i< N}e_{i,i+1}\otimes b_i(n)+z^{-2}e_{N,1}\otimes b_N(n),\
[n]_2=1;\]
\beq\label{Ln}
L_{n}^-(z)=
\sum_{1\le i\le N}e_{i,i}\otimes a^{-1}_i(n)+
\sum_{1\le i< N}e_{i+1,i}\otimes b_i(n)+z^{2}e_{1,N}\otimes b_N(n),\ [n]_2=0.
\eeq
Let $C^*$ be the group of all nonzero complex numbers with respect to
the multiplication. Consider the following action of ${C^*}^{\times 2L}$
on the algebra  $A_q(N,L)$
\[
a_i(n)\mapsto\a_i(n)\ a_i(n)\ \a_i^{-1}(n-1);\]
\[ b_i(n)\mapsto
    \a_i(n)\ b_i(n)\ \a_{i+1}^{-1}(n-1),\quad [n]_2=1;\]
\beq \label{gaugetr}b_i(n)\mapsto \a_{i+1}(n-1)\ b_i(n)\ \a_i^{-1}(n),\quad
[n]_2=0.
\eeq

This action can be represented by the following `gauge
transformation' of matrices $L^{\pm}_n$:
\begin{eqnarray}\label{gaugetr1}
L_{n}^+(z)&\mapsto& D_{n}\ L_{n}^+(z) \ D_{n-1}^{-1},\quad [n]_2=1;\nonumber\\
L_{n}^-(z)&\mapsto& D_{n-1}\ L_{n}^-(z) \ D_{n}^{-1},\quad [n]_2=0,
\end{eqnarray}
where $D_n\in \mbox{End}(C^N)\otimes A_q(N,L)$ act trivially in $A_q(N,L)$ and
diagonally in $C^N$:
\[D_n=\sum_{i=1}^N\alpha_i(n)e_{ii}\otimes1.\]
\begin{prop}
The map:
\[
\chi_i(n)\mapsto  a_{i+1}(n)b_i(n)b_i(n+1)a_i^{-1}(n+1),\quad [n]_2=0;\]
\beq\chi_i(n)\mapsto  a_i(n+1)b_i(n+1)b_i(n)a_{i+1}^{-1}(n),\quad [n]_2=1,
\eeq
extended by linearity to the algebra $C_q(N,L)$, gives a
homomorphism of algebras with the image in the subalgebra
$A^{inv}_q(N,L)\subset
A_q(N,L)$ of elements invariant with respect to the
action of the `gauge group' ${C^*}^{\times 2L}$.
\end{prop}
Denote the images of elements of $C_q(N,L)$ in $A^{inv}_q(N,L)$ by the same
symbols. Clearly, the elements
\[
A_i\equiv\prod_{n=1}^{2L}a_i(n),\ i=1,\ldots,N;\]
\beq
B_k\equiv\prod_{j=1}^Jb_j(2k-2j+1)b_{j+1}^{-1}(2k-2j),\ k=1,\ldots,F,
\eeq
 also belong to $A^{inv}_q(N,L)$. Their products
\beq\label{caz}
z_A\equiv\prod_{i=1}^NA_i,\quad z_B\equiv\prod_{k=1}^FB_k
\eeq
lie in the center of $A_q(N,L)$, so we can consider a representation, where
they act as complex numbers.
\begin{prop}\label{inv}
The  subalgebra $A^{inv}_q(N,L)$ of gauge invariant elements is generated by
the elements $A_1$, $B_1$, and
$\chi_i(n)$, $i=1,\ldots,N$; $n=1,\ldots,2L$,  subject
to the constraints
\[(A_1)^N=z_Af_A(\chi),\quad (B_1)^F=z_Bf_B(\chi),\]
where $f_A(\chi)$ and $f_B(\chi)$ are some monomials of their arguments.
\end{prop}
The proof easily follows from the fact that the ratios $A_iA_{i-1}^{-1}$ and
$B_kB_{k-1}^{-1}$ are monomials
in elements $\chi_j(m)$.

Thus, an appropriate extension of $C_q(N,L)$ by $f^{1/N}_A(\chi)$ and
$f^{1/F}_B(\chi)$ is isomorphic to $A^{inv}_q(N,L)$. Correspondingly, the
evolution (\ref{toda}) can be considered in $A^{inv}_q(N,L)$.

For the description of the integrals of motion we will need also to extend
somehow
the evolution (\ref{toda}) to the bigger algebra $A_q(N,L)$.
One way to do this is to realize the automorphisms $\kappa_\pm$ (\ref{aut}) of
the subalgebra $C_q(N,L)$ by inner ones in $A_q(N,L)$, and then, to extend the
latter to the whole algebra $A_q(N,L)$. There is, however, another
possibility, which provides us with the extension
only up to gauge transformations in the following sense.

Consider a sequence of operators $a_i(n,t),\ b_i(n,t)$, where the `time' index
takes integer values. For each fixed $t$ these elements satisfy the defining
relations of
the algebra $A_q(N,L)$ (\ref{a,b-alg}), (\ref{anbn}). Define the corresponding
elements
of $\mbox{End}(C^N)\otimes A_q(N,L)$ according to (\ref{Ln}):
\beq\label{timeLn}
L^-_{n,t}(z),\ [n+t]_2=0;\quad L^+_{n,t}(z),\ [n+t]_2=1,\eeq
and postulate the zero curvature equation:
\beq
\label{zero}
L^-_{n,t+1}(z)L^+_{n,t}(z)=L^+_{n-1,t+1}(z)L^-_{n-1,t}(z).
\eeq
This equation is invariant with respect to the `time dependent' gauge
transformations:
\begin{eqnarray}\label{gaugetr2}
L_{n,t}^+(z)&\mapsto& D_{n,t}\ L_{n,t}^+(z) \ D_{n-1,t-1}^{-1},\quad
[n+t]_2=1;\nonumber\\
L_{n,t}^-(z)&\mapsto& D_{n-1,t}\ L_{n,t}^-(z) \ D_{n,t-1}^{-1},\quad [n+t]_2=0,
\end{eqnarray}
where
\[D_{n,t}=\sum_{i=1}^N\alpha_i(n,t)e_{ii}\otimes1,\quad \alpha_i(n,t)\in C^*.\]
Like in usual gauge theories, equation (\ref{zero}) does not specify a unique
time evolution in the algebra\footnote{To specify the evolution in gauge
theories, a `gauge fixing procedure' is needed, which is not unique, see e.g.
\cite{SlFa}.} $A_q(N,L)$. The following proposition establishes a direct
relationship of (\ref{zero}) with the
discrete Toda field equations of motion (\ref{toda}).
\begin{prop}\label{prop1}
Let $L^\pm_{n,t}(z)$, defined in (\ref{timeLn}), (\ref{Ln}), be some solution
to equation (\ref{zero}). Then, the gauge invariant (with respect to gauge
transformations (\ref{gaugetr2})) operators
\[
\chi_i(n,t)= a_{i+1}(n,t)b_i(n,t)b_i(n+1,t)a_i^{-1}(n+1,t),\quad
[t]_2-[n]_2=1;\]
\beq\chi_i(n,t)= a_i(n+1,t)b_i(n+1,t)b_i(n,t)a_{i+1}^{-1}(n,t),\quad
[n]_2-[t]_2=1,
\eeq
solve the discrete Toda field equations (\ref{toda}).
\end{prop}
Now, we turn to the description of integrals of motion for the evolution
(\ref{toda}), considered in $A^{inv}_q(N,L)$.

Consider the following `transfer' matrix, element of $A_q(N,L)$:
\beq\label{transfermat}
t_1(z)=\mbox{tr}_{C^N}\left((L^-_{2L}(z))^{-1}L^+_{2L-1}(z)\cdots
(L^-_{2}(z))^{-1}L^+_{1}(z)\right)\in A_q(N,L).
\eeq
In the same way
introduce the set of transfer matrices $t_l(z)$, where $l=2,\ldots,N-1$,
which correspond to elements $L^{\w_l,\pm}(z)\in \mbox{End}(V(\w_l))\otimes
A_q(N)$, where $V(\w_l)$, is the set of irreducible finite dimensional
$U_q(sl(N))$ modules, associated with fundamental weights $\w_l$. These
elements can be constructed through the fusion procedure \cite{KSR} from
$L^{\pm}(z)$.
\begin{prop}
The elements $t_l(z)$, $l=1,\ldots,N-1$, belong to $A^{inv}_q(N,L)$; commute
among themselves
\[ [t_l(z),t_{l'}(z')]=0;\] and  are conserved under the evolution
(\ref{toda}),
extended to $A^{inv}_q(N,L)$.
\end{prop}
These are direct consequences of the gauge transformation law (\ref{gaugetr1}),
the Yang-Baxter equations (\ref{rll}), Proposition \ref{prop1}, and the fusion
procedure \cite{KSR}.

\section{Discrete Toda field theory as a 3-dimentional system}\label{sec3}
\setcounter{equation}{0}

Define the algebra $T_q(N,L)$,
generated by invertible elements $\tau(x,y)$, $x,y\in Z$,
with the following commutation relations
\[
\tau(x,y)\tau(x',y')=\tau(x',y')\tau(x,y),\quad[x+y]_2=[x'+y']_2;\]
\beq\tau(x,y)\tau(x',y')=q^{G(x-x',y-y')}\tau(x',y')\tau(x,y),
\quad [x'+y']_2-[x+y]_2=1,
\eeq
where
\begin{eqnarray}
\lefteqn{G(x,y)=\frac{1}{2}\sum_{j=1}^{N/F}\left(\frac{x-y}{J}+
\varepsilon_J\left(\frac{x-y-1}{2}+jL\right)\right)}\qquad\qquad\nonumber\\
&\times&\left(\frac{x+y}{N}+
\varepsilon_N\left(\frac{x+y-1}{2}+jL\right)\right).,
\end{eqnarray}
Here we used notations (\ref{[n]}) and (\ref{eps_m}).

The function $G(x,y)$ satisfies the following identities:
\[G(x,y)=G(-x,-y),\]
\beq
G(x+1,y)+G(x-1,y)-G(x,y+1)-G(x,y-1)=
2\delta_N(y)\delta_{2L}(x-y).
\eeq
This is the `non-periodic' analog of $g(x,y)$ defined in (\ref{gxy}). Their
difference reads:
\[
G(x,y)-g(x,y)=
\frac{x+y}{2N}\ \varepsilon_L\left(\frac{x-y-1}{2}\right)
+\frac{x-y}{2J}\ \varepsilon_F\left(\frac{x+y-1}{2}\right)
+\frac{x^2-y^2}{2LN}.\]
Define the elements
\beq
\eta(x,y)=(\tau(x-1,y)\tau(x+1,y))^{-1}\tau(x,y-1)\tau(x,y+1).
\eeq
\begin{prop}
The elements
\[\eta(x+2L,y)(\eta(x,y))^{-1},\quad\eta(x+N,y+N)(\eta(x,y))^{-1}\]
lie in the center of the algebra $T_q(N,L)$.
\end{prop}
Using this fact, impose the periodicity conditions:
\beq
\eta(x+2L,y)=\eta(x+N,y+N)=\eta(x,y).
\eeq
Next, consider the following maps of generating elements

\[ \tilde\kappa_\pm\colon\tau(x\mp1,y)\mapsto \tau(x,y),\quad [x+y]_2=0;\]
\begin{eqnarray}\label{map2}
\lefteqn{\tilde\kappa_\pm\colon\tau(x\mp1,y)\mapsto
-(\tau(x+1,y)\tau(x-1,y)}\nonumber\\
&&+\tau(x,y+1)\tau(x,y-1))\tau^{-1}(x,y),\quad[x+y]_2=1.
\end{eqnarray}
\begin{prop}
The maps $\tilde\kappa_\pm$ can be extended by linearity to algebra
automorphisms
\[
\tilde\kappa_\pm\colon T_q(N,L)\rightarrow T_q(N,L),
\]
such that
\beq\tilde\kappa\equiv\tilde\kappa_+\circ\tilde\kappa_-=
\tilde\kappa_-\circ\tilde\kappa_+.
\eeq
\end{prop}
Define now the `time' dependent fields:
\beq
\tilde\kappa^t(\tau(x,y))=\left\{\begin{array}{ll}
                    \tau(2t,x,y),&\mbox{if $[x+y]_2=0$;}\\
                    \tau(2t-1,x,y),&\mbox{otherwise.}
                            \end{array}
                    \right.
\eeq
Notice that the field $\tau(t,x,y)$ is defined only for $[t+x+y]_2=0$.
\begin{prop}
The operators $\tau(t,x,y)$ satisfy the following equations
\beq
\label{Hir}
\tau(t+1,x,y)\tau(t-1,x,y)+\tau(t,x+1,y)\tau(t,x-1,y)+\tau(t,x,y+1)\tau(t,x,y-1)
=0,
\eeq
and permutation relations
\[
\tau(t,x,y)\tau(t-1,x',y')=q^{G(x-x',y-y')}\tau(t-1,x',y')\tau(t,x,y),
\]
\beq
\label{tau}
\tau(t,x,y)\tau(t,x',y')=\tau(t,x',y')\tau(t,x,y).
\eeq
\end{prop}
Equations (\ref{Hir}) were first written by Hirota in \cite{Hirota} for the
scalar `$\tau$-function' $\tau(t,x,y)$, so it is natural to call them as
the `quantum' Hirota equations on `quantum $\tau$-function' (\ref{tau}).

\begin{prop}
There exists an algebra homomorphism
\[\iota\colon C_q(N,L)\ni\chi_i(n)\mapsto-\eta(n+i,i)\in T_q(N,L),\]
such that \[\iota\circ\kappa_\pm=\tilde\kappa_\pm\circ\iota.\]
\end{prop}
Thus, in the `$\tau$'-variables the discrete Toda system can be interpreted as
a 3-dimensional discrete system with equations of motion (\ref{Hir}), which are
invariant with respect to permutations of the space-time coordinates.

\section{The $N\to\infty$ limit as a 3-dimensional thermodynamical limit}
\label{sec4}
\setcounter{equation}{0}

Let us return to the $1+1$-dimensional Toda field theory in the continuous
space-time and consider it as a $2+1$-dimensional field theory in partly
discrete
space-time: one space coordinate is discrete with the values $1,\dots,N$
( the `Lie algebra direction'), the other two are, as usual, continuous.
We will refer to excitations in $1+1$ dimensional theory as
one-dimensional particles and to the excitations in the
$2+1$-dimensional theory as two-dimensional particles.

The spectrum of the theory can be interpreted in two ways: as the
spectrum of the $1+1$-dimensional model and as the spectrum of the
$2+1$-dimensional model:
\begin{itemize}
\item the $1+1$-dimensional interpretation: $N-1$ massive particles with masses
   $M_l=M\sin({{{\pi}l}/{N}})$.
\item the $2+1$-dimensional interpretation: one scalar massless
   particle with the
   momentum in the second space direction given by the above formula.
\end{itemize}

These two interpretations are very reminiscent to the mechanism
of the mass generation via the compactification of extra dimensions
(see for example \cite{GrSchWit}).

As $N\to\infty$ the 2-dimensional momentum of the $2+1$-dimensional scalar
particle becomes continuous. This corresponds to the limit where the ratio
${{\pi l}/{N}}$ in the formula for the masses is kept finite as
$N\to\infty$.
For finite $N$ the scattering of massive one-dimensional particles
is pure elastic with the scattering amplitudes given by (\ref{S-matr}).
The notion of scattering becomes more subtle in the limit  $N\to\infty$,
since we are dealing now with massless $2+1$-dimensional particles.
It is not difficult, however, to verify that the system has the correct
2-dimensional
thermodynamical behaviour in this limit and, to compute the
asymptotics of the free energy (by the 2-dimensional thermodynamical
limit we mean the limit, where the 2-dimensional space volume
of the system increases proportionally to the number of excitations).
It is convenient to use the
thermodynamical Bethe ansatz for these purposes.

\subsection {} The idea of using scattering amplitudes and dispersions
of physical excitations for the description of states
of the thermodynamical equilibrium goes back to \cite{YangYang}.
For the Toda system this gives the
following answer for the energy levels of the model
in the box of length $L$ with periodic boundary conditions
(we assume that $N$ is yet finite):
\beq \label{enfin}
E=\sum_{l=1}^{N}\sum_{\a=1}^{n_l}M\sin({{{\pi}l}\over{N}})\cosh
{\theta_{\a}^{(l)}},
\eeq
where $n_l$ is the number of particles of type $l$ in the state and
the rapidities $\theta_{\a}^{(l)}$ are `quantized' by the periodic
boundary conditions as follows:
\beq \label{bethep}
LM\sin({{{\pi}l}\over{N}})\sinh{\theta_{\a}^{(l)}}=2\pi I_{\a}^{(l)}+
\sum_{(k,\beta)\ne (l,\a)}\phi_{l,k}(\theta_{\a}^{(l)}-
\theta_{\beta}^{(k)}).
\eeq
Here the numbers $I_{\a}^{(l)}$ are integers and $\phi_{l,k}(\theta)=
-i\ln(S_{l,k}(\theta))$ assuming that the branch of the logarithm is chosen
in such a way that $\phi$ vanishes when $b=0$.
Similar equations have been studied in detail in \cite{AW,Wieg}
for the chiral Gross-Neveu type models.

Since the numbers  $I_{\a}^{(l)}$ form only a subset among all
integers, one can introduce the rapidities of `holes' (in the distribution
of  $I_{\a}^{(l)}$ among integers) as solutions to the system
\beq   \label{betheh}
LM\sin({{{\pi}l}\over{N}})\sinh{\tilde\theta_{\a}^{(l)}}=2\pi
\tilde I_{\a}^{(l)}+
\sum_{(k,\beta)}\phi_{l,k}(\tilde\theta_{\a}^{(l)}-
\theta_{\beta}^{(k)}).
\eeq
Here $\tilde I_{\a}^{(l)}$ are all integers which do not
belong to  $\{I_{\a}^{(l)}\}$.

According to \cite{YangYang} we will refer to such a state
as a state with particles with rapidities $\theta$ and with
holes with rapidities $\tilde\theta$.

The 1-dimensional thermodynamical (macroscopic)
states correspond to the limit $L\to\infty$
where $n_l=L\rho_l$ with finite densities $\rho_l$. These states
are parametrized by the asymptotic densities of distributions of
rapidities of particles and holes along the real line ($\rho_l(\theta)$
and $\rho_l^h(\theta)$, respectively). These densities are certainly not
independent. The equations (\ref{bethep}), (\ref{betheh})
provide the following integral equation which relates  $\rho_l(\theta)$
and $\rho_l^h(\theta)$
\beq  \label{eq-dens}
M_l\cosh{\theta}=2\pi\rho_l(\theta)+2\pi\rho_l^h(\theta)+ \sum_{1\le k\le  N}
\int_{-\infty}^{\infty}\phi_{l,k}'(\theta-\a)\rho_k(\a)d\a.
\eeq
The energy of such state grows proportionally to the length of the system:
\beq \label{th-en}
E=L\sum_{1\le l\le N}\int_{-\infty}^{\infty}M_l\cosh{\theta}\rho_l(\theta)
d\theta.
\eeq
The state of the thermodynamical equilibrium minimizes the free energy of
the system which is the linear combination of the energy and the entropy
\beq  \label{fr-en}
F=E-TS.
\eeq
Here $T$ is the temperature and $S$ is the combinatorial entropy of
the gas of particles and holes. It has the following asymptotics
as $L\to\infty$
on macroscopic states:
\begin{eqnarray}  \label{th-entr}
\lefteqn{S=L\sum_{1\le l\le
N}\int_{-\infty}^{\infty}\{(\rho_l(\theta)+\rho_l^h(\theta))
\ln(\rho_l(\theta)+\rho_l^h(\theta))}\nonumber \\
& - &\rho_l(\theta)\ln\rho_l(\theta)
-\rho_l^h(\theta)\ln\rho_l^h(\theta)\}d\theta.
\end{eqnarray}
Minimization of the functional (\ref{fr-en}) with the condition (\ref{eq-dens})
gives the following formula for the free
energy of the state of the thermodynamical equilibrium:
\beq
F=L\sum_{1\le k\le N}\int_{-\infty}^{\infty}M_l\cosh{\theta}\ln(1+
\exp(-{{\e_k(\theta)}/ T}))d\theta,
\eeq
where the functions $\e_k(\theta)$ satisfy the following system of nonlinear
integral equations:
\beq
M_l\cosh{\theta}=\e_l(\theta)+ {1\over{2\pi}}\sum_{1\le k\le  N}
\int_{-\infty}^{\infty}\phi_{l,k}'(\theta-\a)\ln(1+
\exp(-{{\e_k(\a)}/ T})) d\a.
\eeq
\subsection{} Now let us consider the thermodynamical limit and
thermodynamical states of the Toda field theory, regarded as a
$2+1$-dimensional model. It is not difficult to verify that the limit
$N,L\to\infty$ does not
depend on the order in which it is taken. Let us consider the case where
we first take the limit $L\to\infty$ and then $N\to\infty$. The limit
$L\to\infty$ for fixed $N$ has been already described above. When
$N\to\infty$ the 2-dimensional macroscopic states correspond to the macroscopic
number of $2+1$-dimensional excitations. This means that
we have to consider the states with $n=\sigma N$ where
$n=\sum_{1\le l\le N} n_l$ and $\sigma$ is fixed when $N\to\infty$.
The densities of holes and particles in such states will be functions
of 2 variables (of 2-momentum): $\rho_l^h(\theta)\to \rho(\theta,
{{\pi l}/{N}})$, $\rho_l(\theta)\to \rho(\theta,
{{\pi l}/{N}})$.

Let $x={{\pi l}/{N}}$,$y={{\pi k}/{N}}$, and $b$ in (\ref{S-matr})
be fixed, and $N\to\infty$. The function $\phi_{l,k}(\theta)$ has the
following asymptotics in this limit
\beq  \label{ass}
\phi_{l,k}(\theta)={8{\pi}^3(B^2-B)\over N}K(\theta|x,y)+O({1\over N^3}),
\eeq
where the  $B=(1+4\pi/\b^2)^{-1}$ and the function
$K(\theta |x,y)$ has the following form:
\beq  \label{ker}
K(\theta|x,y)=\sinh(\theta)\{{1\over
{\cosh(\theta)-\cos(x+y)}}-{1\over {\cosh(\theta)-\cos(x-y)}}\}.
\eeq
Using asymptotics (\ref{ass}) in equations (\ref{eq-dens}), (\ref{th-en}),
(\ref{th-entr}), we obtain the following description of macroscopic
states in the affine Toda field theory regarded as a $2+1$-dimensional field
theory.

The energy and the entropy of such states are:
\beq \label{the-en}
E={LN\over{\pi}}\int_0^{\pi}\int_{-\infty}^{\infty}M\sin(x)
\cosh({\theta})\rho(\theta,x)d\theta dx,
\eeq
\begin{eqnarray} \label{the-entr}
S&=&{LN\over{\pi}}\int_0^{\pi}\int_{-\infty}^{\infty}
\{(\rho(\theta,x)+\rho^h(\theta,x))
\ln(\rho(\theta,x)+\rho^h(\theta,x))-  \nonumber \\
& &\rho(\theta,x)\ln\rho(\theta,x)
-\rho^h(\theta,x)\ln\rho^h(\theta,x)\}d\theta dx.
\end{eqnarray}
The densities of holes and particles are related by the equation
\begin{eqnarray}  \label{equ-dens}
\lefteqn{M\sin x\cosh{\theta}=2\pi\rho(\theta,x)+2\pi\rho^h(\theta,x)
+8{\pi}^4(B^2-B)} \nonumber \\&\times &\int_0^{\pi}
\int_{-\infty}^{\infty}K'(\theta-\a|x,y)\rho(\a,x)d\a dx.
\end{eqnarray}
Minimizing the free energy (\ref{fr-en}), we obtain the following
expression for the free energy of the Toda model for large $N$:
\beq \label{3den}
F(T)= {LN\over{\pi}}\int_0^{\pi}\int_{-\infty}^{\infty}M\sin(x)
\cosh{\theta}\ln(1+
\exp(-\e(\theta,x)/ T))d\theta dx,
\eeq
where the function $\e(\theta,x)$ is the solution to the following
nonlinear integral equation:
\begin{eqnarray} \label{3dtherm}
\lefteqn{M\sin x \cosh{\theta}=\e(\theta,x)
 + 4{\pi}^3(B^2-B)}\qquad\qquad    \nonumber \\
&\times&\int_0^{\infty}
\int_{-\infty}^{\infty}K'(\theta-\a|x,y)\ln(1+
\exp(-{{\e(\a,y)}/ T})) d\a dy .
\end{eqnarray}
These equations describe the equilibrium thermodynamics of the
Toda system at $N\to\infty$.

\section{Conclusion}
\setcounter{equation}{0}

In this paper we studied the Toda field theory along two lines.
We investigated the two dimensional thermodynamical limit of
this model, and constructed the discrete space-time approximation
which partly has the discrete Lorentz invariance.

\subsection{} Let us analyze the continuum limit in the Toda field theory,
where the third dimension becomes continuous as well. It is not
difficult to see from the Lagrangian (\ref{Toda}) that
such limit corresponds to $\beta\to\infty$ and $M=m\beta$
with $m$ being fixed. As a result we have the theory in $2+1$-dimensional
space-time with the Lagrangian
\begin{equation}  \label{Toda3}
L_{AT}={1\over2}\int \int  \left(({\partial \phi \over \partial t})^2-
({\partial \phi \over \partial x})^2
- 2m^2\exp\left({\partial \phi \over \partial y}\right)\right) dx dy.
\end{equation}
Equations (\ref{3den}) and (\ref{3dtherm}) imply
the following asymptotics of the free energy in this limit:
$$
F(T)=LNT \int_0^{\pi}\int_{-\infty}^{\infty}mt\cosh(\theta)
\ln(1+\exp(-{mt\cosh(\theta)/ T})) d\theta dt.
$$
This is the free energy of massless free particles. Such a behaviour of
the free energy suggests that
in the continuum limit the Toda system describes noninteracting particles.
{}From the structure of the asymptotics of the free energy one can
assume that the particles are fermions.

The fact that in the continuum limit the Toda field theory
describes noninteracting free particles also can be seen from
the corresponding limit in the Bethe equations (\ref{bethep}).
Recall that these equations describe possible values of rapidities of
physical particles in the box of length $L$ with periodic
boundary conditions.
In the continuum limit $\beta\to\infty$, $M=m\beta$, $N=L_1\beta$
with fixed $m$ and $L_1$ the equations (\ref{bethep}) degenerate
into the equations
$$
Lp=2\pi I, \quad
L_1p_1=\pi l,
$$
were $p=mp_1\sinh(\theta)$. The energy  of such excitation, according to
(\ref{enfin}), is
$$
E^2=p^2+p_1^2.
$$
It is clear that this is the spectrum of free relativistic massless
two dimensional particles.
Thus, the Toda theory has some selfinteraction for finite
$\beta$ and $N\to\infty$ but it becomes free in the continuum limit.

\subsection {}
The large $N$ limit in the principal
chiral field theory, based on the group $SU(N)$, has been studied in
the work \cite{PCF}.
The difference between this model and the Toda theory is obvious:
the large $N$ limit of the principal chiral field theory describes
some string-type objects, while the similar limit in the Toda
field theory describes the $2+1$-field theory.

Now let us conclude with some open problems and conjectures.
\begin{itemize}

\item One has to understand the relation between the 3-dimensional
      model constructed in \cite{BaxBaz}, its two dimentional
      counterpart \cite{BKMS}, and the quantum discrete
      system constructed in sections~\ref{sec1}-\ref{sec3}. We conjecture that
      the model constructed in \cite{BaxBaz} corresponds to
      the discrete quantum Toda model at roots of 1.
      The relation
      should be similar to the one between the discrete sine-Gordon
      and the chiral Potts model \cite{BazResh}.

\item When $q$ is a root of 1, the discrete Toda system
      has properties similar to the discrete sine-Gordon system at
      roots of 1: it describes the quantum integrable system,
      interacting with the classical integrable system.

\item It is interesting to compute the spectrum of the
      discrete quantum Toda field theory. By the analogy with the
      continuum model and with the discrete sine-Gordon system, we
      conjecture the following spectrum of the model in the infinite
       interval along the $x$-coordinate.
      Let $T_x$, $T_t$ be the translation operators in $x$ and $t$
      directions, respectively:
$$
T_x\ a_{n,t}\ T_x^{-1}= a_{n+2,t},\ \ T_t\ a_{n,t}\ T_t^{-1}= a_{n,t+2}.
$$
     The common eigenstates of the Hamiltonians, produced by the
     generating function (\ref{transfermat}) and of the operators $T_x$, $T_t$,
     form a Fock space with the same structure of particles as
     in the continuum case, and with  the same scattering amplitudes.
     The translation operators $T_x$, $T_t$ have the following
     eigenvalues on the state with one particle of type $l$
     with the rapidity $\theta$:
$$
T_x|\theta\rangle={\cosh ({\theta \over 2}+\Lambda+{i\pi l\over 2N})
\cosh ({\theta \over 2}-\Lambda+{i\pi l\over 2N})\over
\cosh ({\theta \over 2}+\Lambda-{i\pi l\over 2N})
\cosh ({\theta \over 2}-\Lambda-{i\pi l\over 2N})}|\theta\rangle,
$$
$$
T_t|\theta\rangle={\cosh ({\theta \over 2}+\Lambda+{i\pi l\over 2N})
\cosh ({\theta \over 2}-\Lambda-{i\pi l\over 2N})\over
\cosh ({\theta \over 2}+\Lambda-{i\pi l\over 2N})
\cosh ({\theta \over 2}+\Lambda-{i\pi l\over 2N})}|\theta\rangle.
$$
The eigenvalues of these operators on many particle states are
products over the individual particles of one particle contributions.

\item  The difference between the large $N$ limits in the Toda field
       theories, related to
       other classical Lie algebras and the $SL(N)$,
       can be interpreted as the other (nonperiodic) boundary conditions
       in the extra dimension. It would be interesting to compute the
       corresponding bulk terms in the free energy.

\item We have constructed the quantum analog of the $\tau$-functions,
introduced
      and studied in \cite{Hirota,JiMi}. It would be interesting
      to obtain the formulas for these quantum $\tau$-functions
      which would generalize the  determinant formulas or similar
      constructions, known in the classical case.
\end{itemize}
We are planing to return to these problems in the extended version
of this publication.

{\bf Acknowledgement.}
This work was completed when both of the authors visited
the Laboratoire de Physique Theorique de ENS-Lyon. We are grateful
to the members of the laboratory and especially to Jean-Michel Maillet
and Paul Sorba for the hospitality. The work of R.K. is supported by CNRS.

\end{document}